\documentclass[twocolumn,aps,showpacs,pra,10pt]{revtex4-1}

\usepackage{amsmath}
\usepackage{graphicx}
\usepackage{dsfont}

\newcommand{\im}{{\rm i}}

\newcommand{\ket}[1]{|{#1}\rangle}
\newcommand{\bra}[1]{\langle{#1}|}
\newcommand{\braket}[2]{\langle{#1}|{#2}\rangle}

\begin{document}
\title{High-dimensional quantum channel estimation using classical light}

\author{Chemist M. Mabena}
\affiliation{CSIR National Laser Centre, P.O. Box 395, Pretoria 0001, South Africa}
\affiliation{School of Physics, University of the Witwatersrand, Johannesburg 2000, South Africa}

\author{Filippus S. Roux}
\email{froux@nmisa.org}
\affiliation{National Metrology Institute of South Africa, Meiring Naud{\'e} Road, Brummeria, Pretoria, South Africa}
\affiliation{School of Physics, University of the Witwatersrand, Johannesburg 2000, South Africa}

\begin{abstract}
A method is proposed to characterize a high-dimensional quantum channel with the aid of classical light. It uses a single nonseparable input optical field that contains correlations between spatial modes and wavelength to determine the effect of the channel on the spatial degrees of freedom. The channel estimation process incorporates spontaneous parametric upconversion (sum frequency generation) to perform the necessary measurements.
\end{abstract}

\maketitle

\section{Introduction}

Quantum communication offers a fundamentally secure form of communication \cite{bb84,bb85}. Some of the quantum communication protocols are based on quantum entanglement \cite{e91}. In an ideal world, these protocols offer perfect security, but in practice the channel could introduce noise and distortions that would limit the operation of these schemes.

Different degrees of freedom are used for the photonic implementations of quantum communication. In the paraxial limit, the angular momentum of an electromagnetic field can be decomposed into the spin angular momentum (SAM) and orbital angular momentum (OAM) degrees of freedom \cite{vanenk,oamflux,berrycurr}. An extensive amount of literature has focused on the use of SAM (polarization) \cite{ursin,teleport,satellite,zeiltele2}. However, a disadvantage of polarization is being limited to a two-dimensional Hilbert space. On the other hand, the OAM degrees of freedom define an infinite dimensional Hilbert space \cite{allen,torres,zeil2006,mafu,oamhom,romero}. Such a high-dimensional Hilbert space provides better security and more information capacity \cite{bechmann,cerf,walborn}. Photons can be prepared in an OAM entangled state with the aid of spontaneous parametric down-conversion (SPDC) \cite{mair,miatto2}.

Free-space optical and quantum communication involve transmission through a turbulent atmosphere. If the information or entanglement of the photons is encoded in terms of the spatial (OAM) degrees of freedom, the turbulence, which causes scintillation that distorts the optical field \cite{scintbook}, would result in a loss of information or entanglement \cite{malik,pors,qturb3,oamturb,qkdturb,leonhard,lindb,notrunc,qutrit}. For instance, in a free-space quantum channel, where the entangled state is carried by a pair of photons propagating through turbulence, the distortions due to the scintillation may cause the protocol to fail. The success of such a free-space communication system may thus require the use of an active correction system that compensates for the distortions. Such a correction system involves two parts. One is a channel characterization process that determines the detailed nature of the distortions. The other is a process to remove the distortions based on the information obtained from the characterization process. An example of such a system is an adaptive optical system \cite{aobook2,aobook1,leachao}. It measures the distortions of the wavefront using a wavefront sensor and then removes these distortions with a deformable mirror. For a quantum channel, it is more natural to use quantum process tomography to determine the distortions of the channel \cite{qproctom,qproctom2,nc}. However, the implementation of quantum process tomography is resource intensive \cite{qptomres}. Another approach is to use quantum error correction \cite{qec0,brunqec,toddbrun}.

Recently, a method was proposed to use classical light for the characterization of a quantum channel \cite{chanest}. The spatial degrees of freedom of classical light are affected in the same way as those of single photons when they propagate through turbulence. Therefore, correlations that exist between the spatial degrees of freedom and polarization, which is not affected by turbulence, gives the classical light the same capability as one would have with entangled photons passing through a one-sided channel. As a result, one can exploit the same quantum process tomography formalism to analyze the quantum channel even though one uses classical light for the assessment. It is assumed that the channel retains the purity of the state. When the correlation exists between the spatial modes and polarization, the implementation is limited to two-dimensional spaces (qubits).

The assumption that the channel retains the purity of the state is not a limitation, but rather a benefit. The reason is as follows. A short optical pulse propagating through a turbulent medium, experiences a varying refractive index that is effectively frozen in time. The varying refractive index imparts random phase modulations on the optical field, which means that the scintillation caused by the turbulent medium is a unitary process. Viewed as a quantum channel, it maintains the purity of the state. However, turbulence changes with time. So, if the measurements require an integration period that is longer than the time scale over which the turbulence changes, then it would cause mixing of the state and a loss of purity. It also implies a loss of information, preventing one from being able to correct the channel to the same level as in the unitary case. Therefore, if the scheme allows one to do the characterization of the channel while it maintains purity, one can do a better correction of the channel.

Here, we generalize the classical characterization of quantum channels to higher dimensions by using frequency (wavelength) instead of polarization. As with polarization, frequency is not affected by turbulence; it does not cause cross-talk among different wavelengths. The reason is that turbulence is a linear system that is shift-invariant in time. The amount of scintillation that a turbulent medium imparts on an optical beam propagating through it depends on the wavelength. However, if the bandwidth within which these wavelengths are chosen is small enough, the difference would be negligible.

By implication, we need to replace the polarization optics used in the measurement process with appropriate optical methods that can manipulate the frequency degrees of freedom. To this end, we propose to use spontaneous parametric upconversion (SPUC), also called sum-frequency generation \cite{boydno}. The measurements are made by sending the output after the channel, together with a specially prepared optical field (the measurement `state') through a nonlinear crystal. The idea is that the measurement state will select a specific part of the output state to produce a successful upconversion that would lead to photon detection. In this way, all the different measurements required for the tomography process can be obtained. To show that this scheme works, we first provide a heuristic discussion in Sec.~\ref{idee}. This discussion covers all the aspects of the process, based on assumptions about the nature of the SPUC process. In Sec.~\ref{vol} we then focus on the SPUC process by performing a more detailed calculation to show that the process does indeed work as we propose. We conclude with a few pertinent aspects that are discussed in Sec.~\ref{disc}.

For the spatial degrees of freedom, we'll consider an OAM basis, such as the Laguerre-Gauss (LG) modes \cite{allen}. However, we'll only consider the azimuthal degrees of freedom and ignore the radial degrees of freedom of these modes. It is readily possible to generalize the analysis to incorporate the radial degrees of freedom, but it makes the analysis more complex. The detailed analysis in Sec.~\ref{vol}, uses the LG modes expressed in terms of generating functions, but with the radial index set to zero.

\section{Classical estimation of a quantum channel}
\label{idee}

\subsection{Principle}

The general approach to estimate a quantum channel with the use of classical light assumes that one can use two different degrees of freedom of light, one of which is affected by the channel and another which is not affected. It is also assumed that the channel maintains the purity of any quantum state that propagates through it. One would then prepare an input state (optical field) that contains correlations between these different degrees of freedom. Although a classical optical field, the input field is represented here in terms of Dirac notation
\begin{equation}
\ket{\psi_{\rm in}} = \sum_n \ket{A_n} \ket{B_n} \alpha_n ,
\label{genin}
\end{equation}
where, $A_n$ and $B_n$ represent two different degrees of freedom denoted by separate kets, the $n$ indicates different basis elements in these degrees of freedom and $\alpha_n$ denotes the expansion coefficients, such that $\sum_n |\alpha_n|^2=1$. It is assumed that the channel only affects the $B$ degree of freedom. The output after the channel is obtained by replacing each $\ket{B_n}$ in terms of a superposition of several of them to represent the scattering process.

One can now use tomography measurements to determine the channel. For this purpose, projective measurements are made on both degrees of freedom in terms of the basis given by the original kets and in terms of mutually unbiased basis with respect to these original kets. The benefit of using classical light is that one does not need to do these measurements sequentially. Instead, one can use one bright light pulse with the optical field given by Eq.~(\ref{genin}) and then divide the output into separate channels where all the different measurements are performed simultaneously.

In the previous implementation \cite{chanest}, the $B$ degree of freedom is the spatial modes, which are affected by turbulence in a free-space channel, and the $A$ degree of freedom is polarization, which is not affected by the turbulence. Here, we still have the $B$ degree of freedom as the spatial modes, but the $A$ degree of freedom is the wavelength (frequency), which is also not affected by turbulence.

So, our input state is expressed by
\begin{equation}
\ket{\psi_{\rm in}} = \sum_n \ket{\lambda_n} \ket{\ell_n} \alpha_n
\label{inset}
\end{equation}
where $\lambda_n$ represents the different wavelengths and $\ell_n$ is the azimuthal index of different OAM modes. The channel is a free-space channel and is represented by the effect of turbulence on the OAM modes. The turbulence operation, which is represented by a Kraus operator $\hat{T}$, would scatter the OAM modes to other OAM modes
\begin{eqnarray}
\hat{T} \ket{\ell_n} = \sum_{m} \ket{\ell_m} T_{mn} .
\end{eqnarray}
Here, $T_{mn}$ is a matrix representation of the Kraus operator. It is the purpose of this scheme to determine the matrix $T_{mn}$; to measure the values of its matrix elements. The distortion correction then involves the design of another matrix that would compensate for the distortion caused by the Kraus operator. Hence, after the channel, we obtain
\begin{eqnarray}
\ket{\psi_{\rm out}} = \hat{T} \ket{\psi_{\rm in}} & = & \sum_n \ket{\lambda_n} \hat{T} \ket{\ell_n} \alpha_n \nonumber \\
& = & \sum_{mn} \ket{\lambda_n} \ket{\ell_m} T_{mn} \alpha_n .
\label{uitset}
\end{eqnarray}
Since the indices of $T_{mn}$ are respectively contracted on the basis elements of different degrees of freedom, judicious measurements can uncover all the information (magnitudes and phases) about $T_{mn}$ (apart from an overall phase). Thus, one can recover the Kraus operator for the channel and design another operation that would compensate for it.

When the unaffected degree of freedom is polarization, one can use polarization optics to perform the different projective measurements that are needed to determine the Kraus operator for the channel. In the case where we change the unaffected degree of freedom to be the wavelength, we need a different method to manipulate the wavelength degree of freedom. SPUC provides us with such a method. The measurements are made by preparing optical fields, which we call the measurement states, in terms of combinations of the different bases for the different degrees of freedom. These measurement states are sent through a nonlinear crystal, together with the output field to perform the upconversion.

The main issues concerned with the implementation of this method are: (a) the preparation of the input state; (b) the preparation of the measurement states; and (c) the measurement process itself. We'll start by discussing the preparation of the input state.

\subsection{Input state preparation}

The input state, as generically expressed in Eq.~(\ref{inset}), is by design nonseparable in terms of the two chosen degrees of freedom: wavelength and OAM. To prepare such an optical field, one starts with light consisting of several discrete frequencies, such as a frequency comb laser beam \cite{kam}. The frequency comb laser beam consists of a sequence of laser pulses that occur at a rate given by the pulse repetition frequency. The spectral components of the frequency comb are therefore separated by the pulse repetition frequency, which is typically on the order of a 100~MHz to a few GHz.

\begin{figure}[ht]
\includegraphics{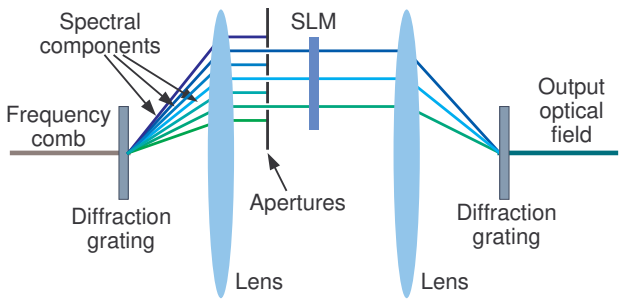}
\caption{Optical setup used to prepare the input optical field.}
\label{inveld}
\end{figure}

The proposed setup to prepare the input optical field is shown in Fig.~\ref{inveld}. One can use a diffraction grating to separate the different spectral components of the frequency comb and some apertures to select out specific components. Note that we select components that are separated by twice the pulse repetition frequency. The reason for this is explained below when we discuss the preparation of the measurement states. With the aid of a lens in a 2-f setup, these components are then incident side-by-side on a spatial light modulator (SLM) where each spectral component is modulated with a specific OAM mode; different OAM modes for different wavelengths.

The spectral components are then recombined using another 2-f system followed by another diffraction grating. The resulting beam now contains correlations between spectral components and the azimuthal indices. In contrast to the vector modes that are used in the previous scheme \cite{chanest}, the current method has no fundamental limitation in the dimensionality of the Hilbert space. Apart from practical limitations such as the size of the SLM, this approach can be used to generate states that are nonseparable in an arbitrary number of dimensions.

The input optical field is sent through the channel to produce the output field, as represented in Eq.~(\ref{uitset}). The Kraus operator only acts on the spatial degrees of freedom and not on the frequency or wavelength degrees of freedom. As a result, the OAM modes are scattered into other OAM modes. Based on the construction of the input state, the output state now contains information about the Kraus operator of the channel. Appropriate measurements on the output state can now uncover the Kraus operator. The measurement process, which employs SPUC, is explained next.

\subsection{Upconversion measurements}

Our proposal is to use the SPUC process to implement the measurement scheme. The inspiration for this proposal comes from studying the spontaneous parametric down-conversion (SPDC) process. In an SPDC process, an input pump beam is incident on a nonlinear crystal. Provided that the phase matching condition is satisfied, there is a probability for a photon from the pump beam to be down-converted to a pair of photons with frequencies that add up to the pump beam frequency. This process incorporates energy and momentum conservation, which implies the creation of entanglement of the respective degrees of freedom in the output photon pair. It has been shown, both theoretically and experimentally, that under general conditions, SPDC conserves OAM \cite{mair}.

For a pump beam with a Gaussian profile, the output after the SPDC process can be expressed as
\begin{equation}
\ket{\Psi_{\rm SPDC}} = \sum_{\ell} \int \ket{\ell,\omega}_s \ket{\bar{\ell},\omega_p - \omega}_i \Lambda(\omega;|\ell|)\ {\rm d} \omega ,
\label{spdc}
\end{equation}
where $\omega_p$ is the angular frequency of the pump, $\ell$ represents the azimuthal index, with $\bar{\ell}=-\ell$, the subscripts $s$ and $i$ denote the signal and idler photons, respectively, and $\Lambda(\omega;|\ell|)$ is a coefficient function.

Ideally, one would like the coefficient function to be constant over a range of azimuthal indices and frequencies. If the frequencies are selected to be relatively close to each other, which is made possible using a frequency comb, one may assume that the coefficient function is independent of frequency over that range. However, it would in general depend on the magnitudes of the azimuthal indices, especially for the high-dimensional case. Based on this understanding, we'll assume that one can replace $\Lambda(\omega;|\ell|)\rightarrow\Lambda(|\ell|)$.

The upconversion process is now given by the dual of the down-converted state $\bra{\Psi_{\rm SPUC}} = (\ket{\Psi_{\rm SPDC}})^{\dag}$, provided that the output that is measured after the upconversion is a photon with a Gaussian profile, analogous to that of the pump beam in the down-conversion process. The tomography measurements are then given by
\begin{equation}
M_{\rm mea} = \braket{\Psi_{\rm SPUC}}{\psi_{\rm out},\psi_{\rm mea}} ,
\label{mmea}
\end{equation}
where $\ket{\psi_{\rm mea}}$ is the measurement state and $\ket{\psi_{\rm out}}$ is the output state after the channel, given in Eq.~(\ref{uitset}). The probability for a successful upconversion (detection of an upconverted photon) is given by $P_{\rm up}=|M_{\rm mea}|^2$. We'll assume the measurement state corresponds to the signal and the output state to the idler, as defined in Eq.~(\ref{spdc}).

\begin{figure}[ht]
\includegraphics{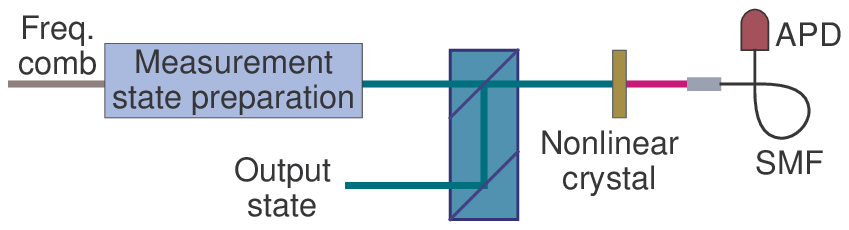}
\caption{Diagrammatic representation of the measurement process.}
\label{meetops}
\end{figure}

One way to look at this process, is to consider the effective state with which the output state is overlapped, by first considering the overlap of the measurement state with the SPUC process,
\begin{equation}
\bra{\psi_{\rm eff}} = \braket{\Psi_{\rm SPUC}}{\psi_{\rm mea}} .
\end{equation}
Consider, for example, the case where the measurement state contains superpositions in both the degrees of freedom
\begin{equation}
\ket{\psi_{\rm mea}} = \left(\ket{\ell_1}a+\ket{\ell_2}b\right) \left(\ket{\omega_1}c+\ket{\omega_2}d\right) ,
\end{equation}
where $a$, $b$, $c$ and $d$ are complex coefficients. The effective state then becomes
\begin{eqnarray}
\bra{\psi_{\rm eff}} & = & \left[\Lambda(|\ell_1|)a\bra{\bar{\ell}_1} + \Lambda(|\ell_2|)b\bra{\bar{\ell}_2}\right] \nonumber \\
& & \times \left(\bra{\omega_p - \omega_1}c+\bra{\omega_p - \omega_2}d\right) .
\label{optoes}
\end{eqnarray}
One can choose the values of $a$ and $b$ to compensate for the difference in magnitude between $\Lambda(|\ell_1|)$ and $\Lambda(|\ell_2|)$. The final measurement is then given by
\begin{equation}
M_{\rm mea} = \braket{\psi_{\rm eff}}{\psi_{\rm out}} .
\end{equation}
We see that $\bra{\psi_{\rm eff}}$ is the dual of a specific state that would be selected by inner product from the output state. Since one can design the measurement state $\ket{\psi_{\rm mea}}$, one has control over $\bra{\psi_{\rm eff}}$.

The measurement can be made with the proposed setup shown in Fig.~\ref{meetops}. The output after the channel is optically combined with a measurement state that is specifically prepared for each measurement. The preparation of the measurement states is discussed in more detail below. The combined beam is then sent through a nonlinear crystal to perform upconversion. The output after the upconversion process is coupled into a single mode optical fiber, which selects out the Gaussian beam profile. The photons that are coupled into the fiber are detected with a single photon detector, such as an avalanche photon diode.

\subsection{Tomography}

To determine the matrix elements of the Kraus operator, one can make specific measurements on the output state, given in Eq.~(\ref{uitset}). One can ignore $\alpha_n$ under the assumption that one can make these values all equal in the preparation of the input state. So, we express the output state as
\begin{equation}
\ket{\psi_{\rm out}} = \sum_{mn} \ket{\ell_m} \ket{\omega_n} T_{mn} ,
\label{uitset0}
\end{equation}
where we converted wavelength into angular frequency.

Clearly, one can obtain the magnitude of every matrix element by using a measurement state with a single frequency and a single OAM mode. The measurement state
\begin{equation}
\ket{\psi_{\rm mea}} = \ket{\bar{\ell}_u}\ket{\omega_p - \omega_v} ,
\end{equation}
where the subscripts $u$ and $v$ represent specific components, leads to
\begin{equation}
\bra{\psi_{\rm eff}} = \bra{\ell_u}\bra{\omega_v} .
\end{equation}
The projective measurement then gives
\begin{eqnarray}
|\braket{\ell_u,\omega_v}{\psi_{\rm out}}|^2 & = & \left| \sum_{mn} \braket{\ell_u,\omega_v}{\ell_m,\omega_n} T_{mn} \right|^2 \nonumber \\
& = & \left| T_{uv} \right|^2 .
\end{eqnarray}

To obtain the relative phases of the elements, one needs to produce superpositions of the frequencies and/or OAM modes. If, for instance, the measurement state contains a superposition of two frequencies
\begin{equation}
\ket{\psi_{\rm mea}} = \ket{\bar{\ell}_u}\left(\ket{\omega_p - \omega_v}+\ket{\omega_p - \omega_w}\right)\frac{1}{\sqrt{2}} ,
\end{equation}
where the subscripts $u$, $v$ and $w$ represent specific components, then we get
\begin{equation}
\bra{\psi_{\rm eff}} = \bra{\ell_u}\left(\bra{\omega_v}+\bra{\omega_w}\right)\frac{1}{\sqrt{2}} .
\end{equation}
In this case, the projective measurement produces an interference term
\begin{eqnarray}
|\braket{\psi_{\rm eff}}{\psi_{\rm out}}|^2 & = &
\frac{1}{2}|\braket{\ell_u,\omega_u}{\psi_{\rm out}}+\braket{\ell_u,\omega_v}{\psi_{\rm out}}|^2  \nonumber \\
& = & \frac{1}{2}\left| T_{uv} + T_{uw} \right|^2 \nonumber \\
& = & \frac{1}{2} \left(\left| T_{uv} \right|^2 + \left| T_{uw} \right|^2 + T_{uv}^* T_{uw} + T_{uv} T_{uw}^*\right) , \nonumber \\
\end{eqnarray}
which can be used to obtain the relative phase of the elements.

\subsection{Measurement state preparation}

\begin{figure}[ht]
\includegraphics{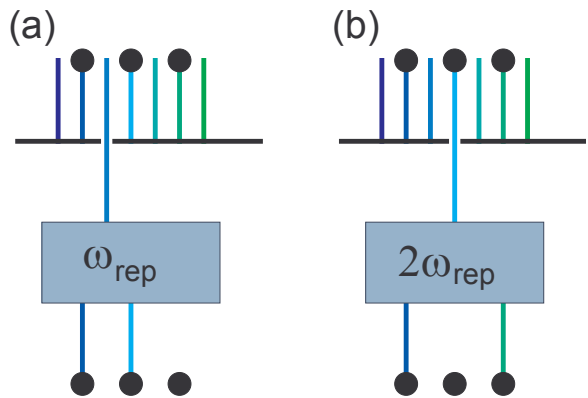}
\caption{Diagrammatic representation of modulation process, using (a) pulse repetition frequency and (b) twice pulse repetition frequency. The black dots denote those spectral components that are chosen as part of the input state.}
\label{ampmod}
\end{figure}

Superpositions of OAM modes are readily produced by the appropriate modulation performed on the SLM. To produce superpositions of frequencies, one can use amplitude modulation with a suppressed carrier. The modulation frequencies need to be on the order of the pulse repetition frequency of the frequency comb. Currently, electro-optical modulators and acousto-optical modulators can perform modulation of optical signals into the GHz range. The modulation processes for two different cases are shown in Fig.~\ref{ampmod}. Remember that the spectral components that are used in the preparation of the input state are separated by twice the pulse repetition frequency.

If the two components in the superposition lie next to each other, separated by twice the pulse repetition frequency, one would select a component between these two components and modulate it by the pulse repetition frequency, as shown in Fig.~\ref{ampmod}(a). This is made possible by the way we have chosen spectral component for the preparation of the input state.

On the other hand, if the spectral components of the superposition are further apart (say four times the pulse repetition frequency), we select a spectral component halfway between the two required components and modulate it by half the separation frequency (two times the pulse repetition frequency), as shown in Fig.~\ref{ampmod}(b).

The relative phase between the two components in the superposition is determined by the phase of the modulation signal. Since multiple measurements with different relative phases need to be made simultaneously, it is necessary to maintain coherence among the different modulation signals in the different measurements.

After the temporal modulation, which produces the required superposition in frequency components, the beam is spatially modulated with a SLM to obtain the spatial profile for the measurement state. This is done by producing a superposition of the spatial mode on the SLM. Depending on the required accuracy of the measurement, the fidelity of the spatial mode can be improved with the aid of complex amplitude modulation \cite{arrizon1}. Note that the measurement state does not have correlations between the bases elements of the different degrees of freedom, as one has for the input state; the measurement states are separable states in terms of these degrees of freedom.

\subsection{Complete process}

\begin{figure}[ht]
\includegraphics{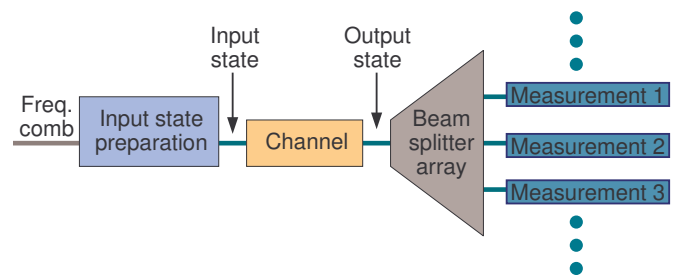}
\caption{Diagrammatic representation of the complete channel estimation process.}
\label{saam}
\end{figure}

Putting the different building blocks together, one obtains an overall process that can be summarized as in Fig.~\ref{saam}. After the input state is prepared, using the process shown in Fig.~\ref{inveld}, and sent through the channel, the output is divided by an array of beam splitters. This allows all the different measurements to be made simultaneously. Each measurement is made with a setup as shown in Fig.~\ref{meetops} and it incorporates the preparation of a measurement state that may require a temporal modulation, as shown in Fig.~\ref{ampmod}.

The information that is obtained from the different measurements is then used to reconstruct the Kraus matrix. This in turn is used to produce a processing matrix that will compensate for the distortions caused by the Kraus operator.

What remains to be done, is to consider the upconversion process in more detail, to confirm the process represented in Eqs.~(\ref{mmea})-(\ref{optoes}). In addition, we need to consider the nature of the coefficient function $\Lambda(|\ell|)$, so that we can understand its effect in Eq.~(\ref{optoes}). For these purposes, we perform a more detailed calculation next.

\section{Explicit calculation}
\label{vol}

Here, we provide a detailed calculation of the upconversion process to show that it does produce the required measurement. This calculation also reveals the nature of the coefficient functions $\Lambda(|\ell|)$ that appear in Eq.~(\ref{spdc}). The calculation can be expressed as an integral representation of Eq.~(\ref{mmea}):
\begin{equation}
M_{\rm mea} = \int \Psi_{\rm SPUC}({\bf a}_1,{\bf a}_2) \psi_{\rm out}({\bf a}_1) \psi_{\rm mea}({\bf a}_2)\ {\rm d}a_1^2\ {\rm d}a_2^2 ,
\label{immea}
\end{equation}
where ${\bf a}_1$ and ${\bf a}_2$ are two-dimensional spatial frequency vectors, associated with the output state and the measurement state, respectively.

To perform the calculation in Eq.~(\ref{immea}) as effectively as possible, we employ some careful considerations and make a few simplifying assumptions. Firstly, we assume that the turbulence in the channel only causes weak scintillation. Under such conditions, the scintillation process can be described by a single phase screen. In turn, this implies that the result is independent of the propagation distance; the propagation has no effect. As a result, both the output state and the measurement state are independent of the wavelength. In the calculation, the wavelengths are determined by those in $\Psi_{\rm SPUC}({\bf a}_1,{\bf a}_2)$.

The SPUC-function $\Psi_{\rm SPUC}({\bf a}_1,{\bf a}_2)$ serves as a kernel function in Eq.~(\ref{immea}). It is composed of the product of the phase matching function $S({\bf a}_1,{\bf a}_2)$ and the mode that couples into the single mode fiber $g({\bf a}_3)$, where ${\bf a}_3$ is the spatial frequency of the upconverted photon. The transverse phase matching conditions replace ${\bf a}_3$ with a linear combination of the spatial frequencies ${\bf a}_1$ and ${\bf a}_2$. As a result, the kernel function for the upconversion process has the form
\begin{equation}
\Psi_{\rm SPUC}({\bf a}_1,{\bf a}_2) = g({\bf a}_1,{\bf a}_2) S({\bf a}_1,{\bf a}_2) .
\label{spucaa}
\end{equation}
The phase matching function is a sinc-function, which one can approximate with a Gaussian function in the limit where the crystal is much shorter than the Rayleigh range of the coupled mode (the thin crystal limit). Hence,
\begin{eqnarray}
S({\bf a}_1,{\bf a}_2) & = & {\rm sinc}\left[ \frac{L}{2}\Delta k_z({\bf a}_1,{\bf a}_2) \right] \nonumber \\
& \approx & \exp\left[ -\frac{L}{2}\Delta k_z({\bf a}_1,{\bf a}_2) \right] ,
\label{faf}
\end{eqnarray}
where $\Delta k_z({\bf a}_1,{\bf a}_2)$ is the mismatch in the longitudinal components of the propagation vectors. Assuming paraxial beams and collinear phase matching, with different wavelengths for the output state and the measurement state, we obtain an expression for $\Delta k_z$, given by
\begin{equation}
\Delta k_z({\bf a}_1,{\bf a}_2) = \frac{\pi n_1 n_2 |\lambda_1 {\bf a}_1-\lambda_2 {\bf a}_2|^2}{\lambda_1 n_2+\lambda_2 n_1} ,
\label{dkz}
\end{equation}
where $\lambda_1$ and $\lambda_2$ are the wavelengths of the output state and the measurement state, respectively, and $n_1$ and $n_2$ are the refractive indices that are, respectively, experienced by the output state and the measurement state in the nonlinear crystal. The transverse phase matching condition gives
\begin{equation}
{\bf a}_3 = \frac{(n_1 {\bf a}_1+n_2 {\bf a}_2)(\lambda_1+\lambda_2) }{\lambda_1 n_2+\lambda_2 n_1} .
\label{a3na12}
\end{equation}
Combining all the above, we obtain an expression for the SPUC kernel-function, given by
\begin{eqnarray}
\Psi_{\rm SPUC}({\bf a}_1,{\bf a}_2) & = & \Omega(\lambda_1,\lambda_2) \exp\left[ - \frac{\pi^2 w_c^2}{n_3^2} |n_1 {\bf a}_1+n_2 {\bf a}_2|^2 \right. \nonumber \\
& & \left. - \frac{\pi n_1 n_2 L |\lambda_1 {\bf a}_1-\lambda_2 {\bf a}_2|^2}{2 (\lambda_1 n_2+\lambda_2 n_1)} \right] ,
\label{spuc0}
\end{eqnarray}
where $w_c$ is the size of the Gaussian mode that couples into the fiber, $\Omega$ is an overall constant that may depend on the wavelengths, and
\begin{equation}
n_3 = \frac{\lambda_1 n_2+\lambda_2 n_1}{\lambda_1+\lambda_2}
\label{n3def}
\end{equation}
is the refractive index seen by the upconverted photon in the nonlinear crystal. Since we are not interested in the global magnitude of the process, as it will cancel out in normalization, we'll henceforth ignore $\Omega$. The expressions in Eqs.~(\ref{dkz}), (\ref{a3na12}) and (\ref{n3def}) are derived in Appen.~\ref{appen1}.

Both the output state and the measurement state can be expressed as superpositions of LG modes. The point of this exercise is to show that the three-way overlap, given in Eq.~(\ref{immea}), serves as an inner product for these LG modes, because this allows the extraction of the coefficients in the expansion from which the information of the Kraus operator is obtained. For this purpose, it suffices to select individual LG modes for the output state and the measurement state. In the calculation, we use the generating function for the angular spectra of the LG modes \cite{ipe} with zero radial index ($p=0$)
\begin{equation}
{\cal G}_{\rm LG}({\bf a};\mu,T,w) = {\cal N} \exp\left[\im \pi (a+\im T b) w \mu - \pi^2 w^2 |{\bf a}|^2\right] ,
\label{lggen}
\end{equation}
where $w$ is the mode size, $T$ is given by the sign of the azimuthal index, $\mu$ is the generating parameter, and
\begin{equation}
{\cal N} = w \left(\frac{2\pi 2^{|\ell|}}{|\ell|!}\right)^{1/2} ,
\label{lgnorm}
\end{equation}
is the normalization constant, which depends on the azimuthal index. The angular spectrum for a particular LG mode is produced by taking the number of derivatives, with respect to the generating parameter equal to the magnitude of the azimuthal index and setting the generating parameter to zero, where after we substitute in the appropriate normalization constant.

One can now substitute Eq.~(\ref{spuc0}) and
\begin{eqnarray}
\psi_{\rm out}({\bf a}_1) & \rightarrow & {\cal G}_{\rm LG}({\bf a}_1;\mu_1,T_1,w_1) \nonumber \\
\psi_{\rm mea}({\bf a}_2) & \rightarrow & {\cal G}_{\rm LG}({\bf a}_2;\mu_2,T_2,w_2) ,
\label{gensubs}
\end{eqnarray}
into Eq.~(\ref{immea}) and evaluate the integrals. We also apply the thin crystal limit, by setting
\begin{equation}
\beta =\frac{L\lambda_3}{\pi w_c^2} \rightarrow 0 ,
\label{betadef}
\end{equation}
where
\begin{equation}
\lambda_3 = \frac{\lambda_1 \lambda_2}{\lambda_1+\lambda_2}
\label{lam3def}
\end{equation}
is the upconverted wavelength. The result is a generating function for the measurements $M_{\rm mea}$, given by
\begin{eqnarray}
{\cal G} & = & \frac{\pi w_1 w_2 {\cal N}_1 {\cal N}_2 n_3^2}{W}\nonumber \\
& & \times \exp\left[ \frac{\mu_1 \mu_2 (1-T_1 T_2) n_1 n_2 w_1 w_2 w_c^2}{2W} \right] ,
\label{mgen}
\end{eqnarray}
where
\begin{equation}
W = n_1^2 w_2^2 w_c^2 + n_2^2 w_1^2 w_c^2 + n_3^2 w_1^2 w_2^2 .
\label{wdef}
\end{equation}

One can use Eq.~(\ref{mgen}) to generate the upconversion probability for any values of the azimuthal indices of the output state and the measurement state. However, it is quite clear that, unless the azimuthal indices have the same magnitude and opposite signs, the result would be zero. It thus follows that the upconversion process does indeed act like an inner product operation (apart from the flip in the sign of the azimuthal index), and that it can therefore be used to extract the information of the Kraus operator from the output state.

The expression in Eq.~(\ref{mgen}) can also be used to determine the nature of the coefficient function $\Lambda(|\ell|)$ that was introduced in Eq.~(\ref{spdc}). Our assumption that $\Lambda(|\ell|)$ is independent of the frequency is validated by the fact that the frequency components in the frequency comb that we select in our implementation, are extremely close to each other. By expanding Eq.~(\ref{mgen}) to all orders in $\mu_1\mu_2$, we obtain the coefficients of the LG modes in the SPDC state (dual of the SPUC process). The coefficients of this expansion are the $\Lambda$-functions and they are given by
\begin{equation}
\Lambda(|\ell|) = \left( \frac{n_1 n_2 w_1 w_2 w_c^2}{W} \right)^{|\ell|} ,
\label{lambf}
\end{equation}
which we have normalized so that $\Lambda(0)=1$. When the measurement state is designed, one can weigh the coefficients of the different LG modes to compensate for the difference in the magnitude of $\Lambda(|\ell|)$, due to different values of $|\ell|$, that would be produced in the upconversion process, as seen in Eq.~(\ref{optoes}).

If the different wavelengths are very close to each other, then the refractive indices seen by the output state and the measurement state would be approximately the same. Critical phase matching would then imply that $n_1=n_2=n_3$. If, furthermore, we assume that the mode sizes of the output state and the measurement state are the same $w_1=w_2=w_0$, then the coefficient function can be expressed as
\begin{equation}
\Lambda(|\ell|) = \left( 2 + \alpha \right)^{-|\ell|} ,
\label{lambf0}
\end{equation}
where
\begin{equation}
\alpha = \frac{w_0^2}{w_c^2} .
\label{alphdef}
\end{equation}
One would prefer that the coefficient function remains as constant as possible as a function of $|\ell|$; in other words, close to 1. From Eq.~(\ref{lambf0}), we see that the best one can do is $\Lambda(|\ell|) = 2^{-|\ell|}$, which implies that the best operating conditions are when $w_0\ll w_c$.

\section{Discussion and conclusions}
\label{disc}

The most common way to characterize a quantum channel is through quantum process tomography. The disadvantage of quantum process tomography is that one needs to prepare a set of basis states. In addition, one also needs to prepare an ensemble of identical input states. In other words, one needs to prepare the same state repeatedly to build up statistics from which one can extract the required information. For an increasing number of dimensions, quantum process tomography quickly becomes unpractically complex. Moreover, due to the required repetition of measurements, the complete measurement process runs over an extended period during which the conditions of the channel may change. As a result, it would not be possible to characterize a single realization of turbulence. The best it can do, is to give information about the average behavior over all turbulence realizations.

In contrast, the use of classical light, as proposed here and in \cite{chanest}, allows one to perform the process with a single pulse of bright light. A higher dimensionality does not lead to longer measurement times, but instead to a more complex measurement system. Therefore, it is not necessary to prepare an ensemble of input states and a large set of input basis states. The single-pulse implementations enable the classical scheme to characterize single realizations of turbulence, which represents a process that maintains purity.

The use of wavelength, instead of polarization as was done in \cite{chanest}, provides the means to implement channel characterization for arbitrary dimensions. There is no fundamental barrier to the number of dimensions. Therefore, the method is scalable. However, the complexity of the measurement system would increase drastically with increasing number of dimensions.

It is important to note that this scheme differs from the conventional classical methods that are based on the preparation of a set of orthogonal input modes, which are used to determine the classical crosstalk matrix. Instead, our scheme only requires the preparation of a single nonseparable input state.

When preparing the input state, one needs to ensure that the modes associated with different wavelength are properly separated on the SLM. To form beams that are well-separated, the incident beam needs to illuminate enough lines on the diffraction grating. The required number of illuminated lines is $N\sim\lambda/\Delta\lambda$. For a pulse repetition frequency of 1~GHZ and a nominal wavelength of $\lambda=1~\mu$m, the number of lines that needs to be illuminated is on the order of 300~000. Diffraction gratings with several thousand lines per mm are commercially available. So, the input beam size needs to be on the order of a 100~mm to illuminate enough lines. Although challenging, this should be doable. Often the best setup for this type of system is to operate the diffraction gratings at the Littrow angle \cite{diffgrat}. This will also avoid too much astigmatism in the first diffracted order.

For a three-dimensional implementation as we show here, the maximum size of the beam on the SLM would be one fifth of the total width of the SLM (taking those beams that are blocked in between, into account). The typical number of pixels across the SLM is on the order of a thousand. Hence, one can expect that each beam on the SLM would have about 200 pixels on a side ($200\times 200$ in total) with which it can be modulated. This should be more than enough to achieve a reasonably well-defined spatial mode. For higher dimensions, one would obtain progressively smaller numbers of pixels. At some point one would need to separate the beams onto different SLMs. This will make the system more complex, but it does not introduce a fundamental limit.

The modulation process that is used in the preparation of the measurement state is assumed to be an amplitude modulation with a suppressed carrier. It should be noted that to implement this form of modulation may require a more involved setup than what a single optical modulator would provide. Details of such a setup would depend on the experimental equipment that are used in the implementation.

In a down-conversion process, the signal and idler photons propagate at particular angles, depending on their frequencies. One can design the down-conversion process so that the signal and idler both propagate collinear with the pump at a particular frequency, but for different frequencies they would have nonzero angles with respect to the pump, because different frequencies are produced with different angles of propagation, as governed by the phase matching conditions of the crystal. The inverse process (upconversion) would therefore require that the different frequencies enter the nonlinear crystal at the required angles to produce a successful upconversion incorporating all the different frequencies. However, if the frequencies are close enough to each other, the differences in these angles would be negligible. This would be the case for a frequency comb where the differences among the different frequency components are on the order of a GHz. In that case, the differences in frequency are too small to produce significantly different propagation angles. (See for instance Appendix C in \cite{shihrep}.)

The upconversion process is assumed to employ critical phase matching, such as type 1 phase matching. For higher efficiency, it is often advisable to use quasi-phase matching that involves type 0 phase matching. This would result in some changes in the details of the calculation. However, these changes are not expected to give significantly different results.

In the practical implementation of the measurements, it is assumed that the wavelength would always match up with a particular wavelength produced in the upconversion process. However, if the different wavelength components, from which the input state and measurements states are composed, are extremely close together, then the upconversion process would produce upconverted wavelengths that are also very close together. As a result, a wavelength filter may not be able to separate them to select the particular wavelength intended in the design. It may therefore require an additional diffraction grating-based filter after the upconversion process to select the specific wavelength that would enable the correct operation of the measurement process.

\section*{Acknowledgements}

We would like to thank Andrew Forbes and Bienvenu Ndagano for illuminating discussions. CMM acknowledges support from the CSIR National Laser Centre.

\appendix

\section{Phase matching conditions}
\label{appen1}

Any deviation from momentum conservation in SPDC can be expressed in terms of the three-dimensional propagation vectors ${\bf k}$ as
\begin{equation}
\Delta {\bf k} = n_1 {\bf k}_1 + n_2 {\bf k}_2 - n_3 {\bf k}_3 ,
\label{dk}
\end{equation}
where the subscripts $1,2,3$ refer to the signal, idler and pump beams, respectively. The refractive indices $n_1$, $n_2$ and $n_3$ depend on the frequencies and states of polarization of their associated optical beams.

For critical phase matching, the dominant part of Eq.~(\ref{dk}) becomes zero. As a result, if we assume collinear conditions and replace the three propagation vectors by their magnitudes, we obtain a relationship among the refractive indices and wavelengths
\begin{equation}
0 = \frac{n_1}{\lambda_1} + \frac{n_2}{\lambda_2} - \frac{n_3}{\lambda_3} .
\label{mag}
\end{equation}
Energy conservation provides us with another relationship among the three wavelengths given in Eq.~(\ref{lam3def}). Using Eqs.~(\ref{mag}) and (\ref{lam3def}), one can obtain an expression for $n_3$, given by
\begin{equation}
n_3 = \frac{n_1\lambda_2 +n_2\lambda_1}{\lambda_1+\lambda_2} .
\label{defn3}
\end{equation}

The transverse part of the mismatch given in Eq.~(\ref{dk}) is zero $\Delta {\bf k}_T=0$, because one can extend the transverse spatial integrations to infinity, provided that the pump beam does not overfill the nonlinear crystal. Here, the subscript $T$ represents the transverse part. As a result, the transverse components of the pump beam can be replaced by
\begin{equation}
{\bf a}_{3} = \frac{n_1 {\bf a}_{1} + n_2 {\bf a}_{2}}{n_3} = \frac{(n_1 {\bf a}_{1} + n_2 {\bf a}_{2})(\lambda_1+\lambda_2)}{n_1\lambda_2 +n_2\lambda_1} ,
\label{dkt}
\end{equation}
where we express the transverse parts of the propagation vectors in terms of two-dimensional spatial frequency vectors, related by $2\pi{\bf a}={\bf k}_T$.

Due to the finite length of the nonlinear crystal, the longitudinal components do not cancel exactly
\begin{equation}
\Delta k_z = n_1 k_{1z} + n_2 k_{2z} - n_3 k_{3z} .
\label{dkza}
\end{equation}
One can express the longitudinal components of the propagation vectors in terms of the transverse spatial frequency vectors
\begin{equation}
k_{mz} = 2\pi\left(\frac{1}{\lambda_m^2}-|{\bf a}_{m}|^2\right)^{1/2} ,
\label{az}
\end{equation}
where $m=1,2,3$. Substituting these expressions into Eq.~(\ref{dkza}), performing a paraxial expansion and applying Eqs.~(\ref{lam3def}) and (\ref{defn3}), we then obtain (ignoring an overall minus sign)
\begin{equation}
\Delta k_z = \frac{\pi n_1 n_2 |\lambda_1{\bf a}_{1}-\lambda_2{\bf a}_{2}|^2}{n_1\lambda_2+n_2\lambda_1} .
\label{dkz0}
\end{equation}


\end{document}